# Theory of nonreciprocal spin waves excitation in spin-Hall oscillators with Dzyaloshinkii-Moriya interaction

R. Zivieri[1], A. Giordano[1], R. Verba[2], B. Azzerboni[3], M. Carpentieri[4], A.N. Slavin[5], G. Finocchio[1,*]

[1]Department of Mathematical and Computer Sciences, Physical Sciences and Earth Sciences, University of Messina, Messina, Italy

[2]Institute of Magnetism, Kyiv, 03680, Ukraine

[3]Department of Engineering, University of Messina, Messina, Italy

[4]Department of Electrical and Information Engineering, Politecnico di Bari, I-70125 Bari, Italy

[5]Department of Physics, Oakland University, Rochester, MI 48309, USA

*gfinocchio@unime.it

**Abstract**

A two-dimensional analytical model for the description of the excitation of nonreciprocal spin waves by spin current in spin-Hall oscillators in the presence of the interfacial Dzyaloshinskii-Moriya interaction ($i$-DMI) is developed. The theory allows one to calculate the threshold current for the excitation of spin waves, as well as the frequencies and spatial profiles of the excited spin wave modes. It is found, that the frequency of the excited spin waves exhibits a quadratic red shift with the $i$-DMI strength. At the same time, in the range of small and moderate values of the $i$-DMI constant, the averaged wave number of the excited spin waves is almost independent of the $i$-DMI, which results in a rather weak dependence on the $i$-DMI of the threshold current of the spin wave excitation. The obtained analytical results are confirmed by the results of micromagnetic simulations.

PACS numbers: 75.30.Ds, 76.50.+g, 75.78.Cd, 75.76.+j

# I. Introduction

In recent years, the excitation of microwave magnetization oscillations driven by a spin-polarized electric current or pure spin current has attracted much attention , both among theoreticians and experimentalists. Magnetization dynamics in spin-torque oscillators (STOs) and spin-Hall oscillators (SHOs) can exhibit various types of behaviour, including highly nonlinear and non-stationary dynamics[1, 2, 3], making these oscillators an  interesting test system for the investigation of nonlinear phenomena in ferromagnets. At the same time, STOs and SHOs demonstrate properties, that make them suitable for a wide range of applications, such as generators of microwave signals[4, 5, 6, 7, 8], neuromorphic computing[9], microwave-assisted magnetic recording[10], etc.

The STOs and SHOs, in which spin-polarized electric current (or pure spin current) is injected locally in an unbounded ferromagnetic layer, are an important class of oscillators[11, 12, 13] , because *propagating* spin waves can be excited in these oscillators  in the case of out-of-plane magnetization [14, 15, 16, 17, 18, 19, 20, 21, 22]. Excitation of *propagating* spin waves makes these oscillators promising for signal-processing applications in all spin-wave logic[23] and magnonics[24]., and  for the development of large arrays of phase-locked auto-oscillators efficiently coupled by  the propagating spin waves [25, 26, 27].

In the case when the SHO free layer is influenced by the interfacial Dzyaloshinskii–Moriya Interaction[28, 29] (*i*-DMI), which is an antisymmetric exchange interaction, appearing at the interface between a ferromagnet and a heavy metal with large spin-orbit coupling[30], the SHO could acquire an additional functionality. The *i*-DMI is known to introduce the frequency nonreciprocity into the spectrum of propagating spin waves[30, 31, 32, 33], leading to several potential physical and technological implications, such as creation of unidirectional spin-wave emitters, separation of signal and idler waves in frequency and wavenumber domains in spin-wave devices, which use parametric and nonlinear spin-wave processes, etc. [34, 35, 36, 37]. In recent theoretical works,[38, 39] it has been shown that the *i*-DMI in STO and SHO results in the excitation of two-dimensional nonreciprocal spin waves, and, at a sufficient strength of the *i*-DMI, in the generation of spiral spin wave modes.

The main purpose of this work is the development of an analytical model, which describes the excitation of *two-dimensional* nonreciprocal spin waves in a nanocontact SHO (the quasi-one-dimensional case of a nanowire-based SHO has been already considered theoretically in Ref. [38]). Our approach is based on an approximate solution of the linearized Landau-Lifshitz-Gilbert-Slonczewski (LLGS) equation and, in fact, is a generalization of the Slonczewski's theory [14] to the case of the presence of the *i*-DMI. The developed theory allows one to calculate profiles of the excited spin waves, which are approximately described by a combination of Laguerre's polynomials and Tricomi's hypergeometric functions, as well as to calculate the excitation threshold and frequency of excited spin waves, which both become lower with the increased *i*-DMI strength.

The paper is organized as follows. Sections II describes the model system used in this study. In Sec. III a step-by-step derivation of the analytical formalism is presented. Analytically calculated results are compared with micromagnetic modeling in Sec. IV. Finally, conclusions are given in Sec. V.

## II. Device under study and micromagnetic simulations

The device under investigation is shown in Fig. 1. It is a typical SHO, consisting of a ferromagnetic/ heavy metal bilayer. The current is injected locally in the bilayer by using a gold concentrator of a double-triangular shape with a distance $d$ between the tips. The system is biased by an external magnetic field $B_{ext}$, applied in the *y-z* plane and making the angle $\theta_B$ with the film normal (axis *z*) (Fig. 1(b)). The bias magnetic field is required in order to tilt the film static magnetization from the in-plane direction, and, if the angle $\theta_M$ between the static magnetization and film normal is sufficiently small, the SHO supports excitation of propagating spin waves.

Otherwise, either a nonlinear self-localized bullet mode is excited due to the negative nonlinear frequency shift, or a transient regime of mode coexistence is realized[17, 40].

In our micromagnetic simulations we used the parameters of a Pt(5nm)/CoFeB(1nm) bilayer, having a rectangular in-plane cross-section of 1500 nm × 3000 nm. The gold concentrator was assumed to be 150 nm thick, with the distance between the tips of $d$ = 100 nm. Details on the calculation of the electric current and the spin current profiles can be found in Ref. [38]. For the materials parameters of the ferromagnetic layer we assumed: gyromagnetic ratio $\gamma = 2\pi \times 28$ GHz/T, the saturation magnetization $M_S = 1000 \times 10^3$ A/m, the exchange stiffness $A = 2.0 \times 10^{-11}$ J/m, the constant of perpendicular surface anisotropy $K_s = 5.5 \times 10^{-4}$ J/m$^2$ (resulting in the effective volume anisotropy of $K_u = 5.5 \times 10^5$ J/m$^3$), the Gilbert damping parameter $\alpha_G = 0.03$ and the spin-Hall angle $\alpha_H = 0.1$. The $i$-DMI parameter was varied in a range [41] in order to systematically study its effect on the nonreciprocal propagation of spin waves. Experimentally a $i$-DMI parameter variation can be realized by the variation of the ferromagnetic film thickness or by use of a different material, covering the ferromagnetic film from another side. The external bias magnetic field was applied at the angle $\theta_B = 15°$. For these parameters, the CoFeB layer had an easy-plane total (material plus shape) anisotropy. It is known, that a partial compensation of the demagnetization field by the perpendicular anisotropy allows one to reduce the critical current density necessary to excite propagating spin wave modes in a tilted external field[38]. All the micromagnetic simulations in this study have been performed using a state of the art micromagnetic solver[42].

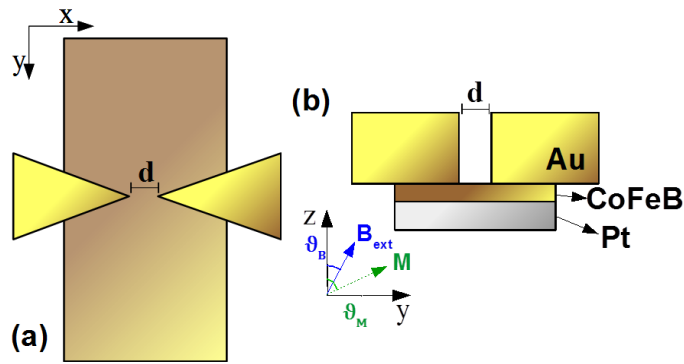

Fig. 1. Sketch of the device under investigation, in-plane view (a) and *x-z* cross-section (b). The direction of the applied field $\boldsymbol{B}_{ext}$ with the indication of the angle $\theta_B$ with respect to the z axis and the angle $\theta_M$ of the equilibrium magnetization $\boldsymbol{M}$ vector is also shown.

## III. Analytical model

**(a) Initial equations**

Dynamics of magnetization $\boldsymbol{M}(\boldsymbol{r},t)$ of a ferromagnetic layer under the influence of spin current is described by the LLGS equation:

$$\frac{d\boldsymbol{M}}{dt} = \gamma \boldsymbol{B}_{eff} \times \boldsymbol{M} + \frac{\alpha_G}{M_s}\boldsymbol{M} \times \frac{d\boldsymbol{M}}{dt} - \frac{g\mu_B \alpha_H}{2eM_S^2 t_{FM}} \boldsymbol{M} \times \boldsymbol{M} \times (\boldsymbol{e}_z \times \boldsymbol{J}) \quad (1)$$

where $g$ is the Landè factor, $\mu_B$ is the Bohr magneton, $e$ is the electron charge, $t_{FM}$ is the thickness of the ferromagnetic layer, $\alpha_H$ is the spin-Hall angle and $\boldsymbol{J}$ is the electric current density flowing in the Pt layer. The effective field $\boldsymbol{B}_{eff}$ includes the contributions of external field $\boldsymbol{B}_{ext}$, demagnetization, exchange, and *i*-DMI contributions ($\boldsymbol{B}_{i-DMI} = 2D/M_s^2 [(\nabla \cdot \boldsymbol{M})\boldsymbol{e}_z - \nabla M_z]$, where $D$ is the *i*-DMI constant).

Equation (1) is used in micromagnetic simulations, but it is too complex for the analytic analysis. From Eq. (1) one can derive a dispersion relation of linear spin waves propagating in the ferromagnetic film (for this purpose one needs to neglect 2 last non-conservative terms and to represent the full magnetization of the film as a sum of its static magnetization and a small dynamic deviation)[31]:

$$\omega_k = \sqrt{(\omega_H + \omega_M \lambda^2 k^2)(\omega_H + \omega_M \lambda^2 k^2 + \omega_M (1-N_{an})\sin^2 \theta_M)} + \omega_M \tilde{D} k_x, \quad (2)$$

where $\boldsymbol{k}$ is the wave vector of a spin wave, $\omega_H = \gamma B_{eff}$, $B_{eff}$ is the effective static magnetic field, $\omega_M = \gamma \mu_0 M_s$  $N_{an} = 2K_u/(\mu_0 M_s^2)$, where $K_u$ is the anisotropy constant, $\lambda = \sqrt{2A/(\mu_0 M_s^2)}$ is the material exchange length, and $\tilde{D} = 2D\sin\theta_M / (\mu_0 M_s^2)$ is the normalized i-DMI constant. Since we consider an ultrathin ferromagnetic film, the in-plane dynamic dipolar contribution is neglected in Eq. (2). In the range $\omega_M \lambda^2 k^2 \ll \omega_0$ the dispersion relation can be approximated as:

$$\omega_k \approx \omega_0 + \omega_M \tilde{\lambda}^2 k^2 + \omega_M \tilde{D} k_x, \qquad (3)$$

where $\omega_0 = \sqrt{\omega_H (\omega_H + \omega_M (1 - N_{an})\sin^2\theta_M)}$ is the ferromagnetic resonance frequency and $\tilde{\lambda}^2 = \lambda^2 (2\omega_H + \omega_M (1 - N_{an})\sin^2\theta_M) / 2\omega_0$.

Making a formal substitution $k_x \to -i(d/dx)$, $k_y \to -i(d/dy)$ in the simplified dispersion equation, it is possible to obtain the following dynamical equation describing the spatial and temporal evolution of the spin wave complex amplitude $a$:

$$\frac{\partial a}{\partial t} = -i\omega a = -i\left(\omega_0 - \omega_M \tilde{\lambda}^2 \nabla^2 - i\omega_M \tilde{D}\frac{\partial}{\partial x}\right)a - \alpha_G \omega a + \sigma J(r) a, \qquad (4)$$

which differs from the one used by Slonczewski[14] by the presence of the i-DMI term. The spin wave damping is accounted for by the term $\alpha_G \omega$, while the influence of the spin current could be easily calculated from Eq. (1) within the framework of the perturbation theory[43], and is given by the term $\sigma J(r) a$ with the coefficient $\sigma = g\mu_B \alpha_H \sin\theta_M / (2eM_s t_{FM})$, describing the spin-Hall efficiency.

We have not included the Oersted field in the model (which results in a spatial dependence of $\omega_0$), because it does not introduce any qualitative change[38]. Thus, the only spatially dependent parameter in Eq. (4) is the distribution of the current density. We approximate it by the function $J(r) = J$ if $r < R_{eff}$ and $J(r) = 0$ otherwise, i.e. assume that current is flowing only within a circle of the radius $R_{eff}$. For spin-Hall oscillators with concentrators like the one shown in Fig. 1 it is an approximation, and the value of the effective radius $R_{eff}$, which is of the order of the half distance between the concentrator tips, should be determined by comparison with simulations (see Sec. IV). Simultaneously, such a case can be exactly realized in an STO[39].

**(b) General solution of the eigenvalue problem**

Equation (4) can be considered as an eigenvalue problem, whose solution gives the values of the spin wave excitation frequency $\omega$ and the critical current $J$. In the considered geometry it is convenient to express Eq. (4) in cylindrical coordinates $(\rho, \phi)$:

$$\left(\frac{\partial^2}{\partial \rho^2} + \frac{1}{\rho}\frac{\partial}{\partial \rho} + \frac{1}{\rho^2}\frac{\partial^2}{\partial \phi^2}\right)a + i\tilde{A}\left(\cos\phi\frac{\partial}{\partial \rho} - \frac{\sin\phi}{\rho}\frac{\partial}{\partial \phi}\right)a + (W + iG)a = 0. \tag{5}$$

Here we introduce a dimensionless coordinate $\rho = r/R_{\text{eff}}$, and the following dimensionless parameters: $\tilde{A} = \tilde{D}R_{\text{eff}}/\tilde{\lambda}^2$, describing the strength of the $i$-DMI, $W = (\omega - \omega_0)R_{\text{eff}}^2/(\omega_M \tilde{\lambda}^2)$, proportional to the generation frequency offset from the FMR frequency, and the normalized total damping $G$, which is equal to $G_1 = (\alpha_G \omega - \sigma J)R_{\text{eff}}^2/(\omega_M \tilde{\lambda}^2)$ within the active region ($\rho < 1$) and to $G_2 = (\alpha_G \omega)R_{\text{eff}}^2/(\omega_M \tilde{\lambda}^2)$ outside the active region.

Equation (5) does not allow an exact analytical solution, because the dependencies on the radial and azimuthal coordinates cannot be separated due to the presence of the $i$-DMI term. At the same time, in the absence of the $i$-DMI this separation can be done rigorously, and the solution, corresponding to the lowest excitation threshold has a simple form $a = a(\rho)$, i.e. it is radially symmetric, and does not depend on the azimuthal angle $\phi$. Hence, we can assume, that, at least in the range of a relatively weak $i$-DMI, the radially symmetric solution is only weakly modified, and the dependence on $\phi$ is also weak. This approximation allows us to consider the azimuthal coordinate not as an independent variable, but as a parameter, which affect the radially symmetric solution $a = a_\phi(\rho)$, i.e. to neglect the derivative $\partial/\partial \phi$ in Eq. (4). As will be shown below, this approximation leads to correct dependencies of the generation frequency and threshold in the i-DMI range of interest.

Owing to the mentioned approximation, Eq. (5) is simplified to:

$$\left(\frac{\partial^2}{\partial \rho^2} + \frac{1}{\rho}\frac{\partial}{\partial \rho} + i\tilde{A}\cos\phi \frac{\partial}{\partial \rho}\right)a + (W + iG)a = 0. \tag{6}$$

Equation (6) is a generalized confluent Riemann hypergeometric equation. Its general solution is a linear combination of a Laguerre's polynomial $L$ (often known as a particular form of a Kummer's hypergeometric function) and a confluent hypergeometric function $U$ (often known as a Tricomi's hypergeometric function) times an exponential function, namely:

$$a_\phi(\rho) = e^{-i(\alpha+\beta)\rho/2}\left[C_1 L\left(-\frac{1}{2} - \frac{\alpha}{2\beta}, i\beta\rho\right) + C_2 U\left(\frac{1}{2} + \frac{\alpha}{2\beta}, 1, i\beta\rho\right)\right], \tag{7}$$

where the parameters $\alpha$ and $\beta$ are defined as: $\alpha = \tilde{A}\cos\phi$ and $\beta_{1,2} = \sqrt{4(W + iG_{1,2}) + \tilde{A}^2\cos^2\phi}$.

The coefficients $C_1$, $C_2$ should be determined from the boundary conditions and proper asymptotes. Since the function $U$ is divergent at $\rho \to 0$, the solution in the active region ($\rho < 1$) is given by the Laguerre's polynomial solely:

$$a_{\phi,1}(\rho) = e^{-i(\alpha+\beta_1)\rho/2} L\left(-\frac{1}{2} - \frac{\alpha}{2\beta_1}, i\beta_1\rho\right). \tag{8}$$

The solution outside the active region should have the asymptotic form of a decaying propagating wave, i.e $a_{\phi,2}(\rho) \sim \rho^{-1/2} e^{i\kappa\rho} e^{-c_g G_2 \rho}$ with $c_g > 0$. This property is satisfied by the following combination:

$$a_{\phi,2}(\rho) = Ce^{-i(\alpha+\beta_2)\rho/2}\left[L\left(-\frac{1}{2} - \frac{\alpha}{2\beta_2}, i\beta_2\rho\right) - \frac{i^{1+\alpha/\beta_2}}{\Gamma[1/2 - \alpha/(2\beta_2)]} U\left(\frac{1}{2} + \frac{\alpha}{2\beta_2}, 1, i\beta_2\rho\right)\right], \tag{9}$$

where $\Gamma[x]$ is the gamma-function. The coefficient $C$ is determined by the continuity of the solution at the boundary of the active region: $a_{\phi,1}(1) = a_{\phi,2}(1)$. In the case of zero i-DMI, $\alpha = 0$, the above solutions are simplified to $a_1(\rho) = J_0(\beta_1 \rho / 2)$ and $a_2(\rho) = CH_0^{(1)}(\beta_2 \rho / 2)/2$, respectively, where $J_0$ and $H_0^{(1)}$ are the Bessel and Hankel functions of the zero order, which is in full accordance with Ref. 14. [44]

**(c) Angular dependence of spin-wave wave number**

Using asymptotic expansions of Laguerre polynomial and hypergeometric function one can show, that at $\rho \gg 1$ the solution of Eq. (9) behaves as $a_{\phi,2}(\rho) \sim \rho^{-1/2-\alpha/2\beta_2} \exp[i(\beta_2 - \alpha)\rho/2]$, i.e. has a form of a wave, propagating from a point source, and having an angular-dependent wave number, which is determined by the term $\exp[ik_\phi r]$. The wave number is equal to $k_\phi = \mathrm{Re}[\beta_2 - \alpha]/(2R_{\mathrm{eff}})$, or, in the initial parameters can be expressed as:

$$k_\phi = \frac{1}{2\tilde{\lambda}^2}\left[-\tilde{D}\cos\phi + \sqrt{4\frac{\omega-\omega_0}{\omega_M}\tilde{\lambda}^2 + \tilde{D}^2\cos^2\phi}\right]. \tag{10}$$

This expression can be also directly obtained from the spin-wave spectrum Eq. (3), which confirms the correct asymptotic behavior of the solution given by Eqs. (8, 9). The exponential decay of the spin waves, caused by damping, is described by the term $\exp[-\alpha_G r / v_{gr}]$ with

$$v_{gr} = \omega_M\left(2\tilde{\lambda}^2 k + \tilde{D}\cos\phi\right) \tag{11}$$

being the spin-wave group velocity (to derive this expression we used the assumption of small damping, $\alpha_G \ll 1$).

The dependence of the spin-wave wave number on the azimuthal angle is nonreciprocal, in the sense that $k_\phi \neq k_{\pi-\phi}$, which is a consequence of the i-DMI. The averaged value of the wave number is equal to

$$\langle k \rangle = \frac{\sqrt{4(\omega-\omega_0)\tilde{\lambda}^2/\omega_M + \tilde{D}^2}}{\pi\tilde{\lambda}^2} E\left[\frac{\tilde{D}^2\omega_M}{4(\omega-\omega_0)\tilde{\lambda}^2 + \tilde{D}^2\omega_M}\right], \tag{12}$$

where $E[m]$ is the complete elliptic integral of the second kind. For small *i*-DMI it is simplified to: $\langle k \rangle = \sqrt{4(\omega - \omega_0)\tilde{\lambda}^2 / \omega_M + \tilde{D}^2} / (2\tilde{\lambda}^2)$. In the section below, we will find the excitation frequency $\omega$, and will show, that the averaged value of the spin-wave wave number is almost independent of $\tilde{D}$ in the range of a relatively weak *i*-DMI

**(d) Determination of the threshold current and generation frequency.**

The generation frequency and threshold current density can be determined by the application of the boundary conditions to the general solution Eqs. (8, 9). The boundary conditions require continuity of the function $a_\phi(\rho)$ and its derivative at the boundary of the active region ($\rho = 1$). The first condition is satisfied automatically by the selection of the coefficient $C$ in Eq. (9). However, since we use approximate solutions, the condition on the derivatives $da_{\phi,1}/d\rho|_{\rho=1} = da_{\phi,2}/d\rho|_{\rho=1}$ cannot be satisfied exactly for all the azimuthal angles $\phi$ simultaneously by any values of the generation frequency and the bias current density. Therefore, instead of the condition of the exact matching of derivatives, we use the condition of the minimization of a total mismatch of the derivatives. For this purpose, we construct the functional of the quadratic deviation of the derivatives at the boundary of the active region:

$$\Phi[W, G_1] = \int_0^{2\pi} |\mathcal{F}(\phi)|^2 d\phi, \qquad (13)$$

where

$$\mathcal{F}(\phi) = \left( \frac{da_{\phi,1}}{d\rho} - \frac{da_{\phi,2}}{d\rho} \right)\bigg|_{\rho=1}. \qquad (14)$$

The normalized generation frequency $W$ and the threshold $G_1$ are, then, given by the minimum of $\Phi[W, G_1]$.

Let us find an analytical approximation for the generation frequency and threshold. Taking into account the structure of the functions $a_{\phi,i}(\rho)$, we can consider the function $\mathcal{F}$ as the function of three variables: $\alpha = \tilde{A}\cos\phi$, $\beta_1$ and $\beta_2$. The value of $\alpha$ is proportional to the *i*-DMI strength, which is considered relatively small in the model. Thus, we can expand the function $\mathcal{F}$ in a series leaving only a linear term in $\alpha$, namely $\mathcal{F} = \mathcal{F}_0 + C_f \tilde{A}\cos\phi$, where $\mathcal{F}_0 = \mathcal{F}(\tilde{A}=0)$. After the integration one gets $\Phi = \int_0^{2\pi}|\mathcal{F}_0|d\phi + |C_f|^2 \tilde{A}^2/2$. Consequently, the condition of the function minimum $\partial\Phi/\partial W = \partial\Phi/\partial G_1 = 0$ does not depend on $C_f$. This means, that we can set $\alpha = 0$ in the definition of the function $\mathcal{F}$, at least for a small *i*-DMI. This property is, in fact, more general – the generation frequency and threshold should be the same for *i*-DMI of the same strength but opposite values, because the change $D \to -D$ corresponds to the simple inversion of the *x*-axis. Thus, odd functions of *D* can be safely disregarded.

Setting $\alpha = 0$ the function in Eq. (13) is simplified to:

$$\mathcal{F} = \frac{\beta_1}{2}J_0\left(\frac{\beta_1}{2}\right)H_1^{(1)}\left(\frac{\beta_2}{2}\right) - \frac{\beta_1}{2}J_1\left(\frac{\beta_1}{2}\right)H_0^{(1)}\left(\frac{\beta_2}{2}\right) \tag{15}$$

Following Ref. 14, we, first, consider the case of zero Gilbert damping. Then, the function of Eq. (15) has exact zero at the values $W + \tilde{A}^2\cos^2\phi/4 \approx 1.43$ and $G_1 = -\sigma J R_{\text{eff}}^2/(\omega_M \tilde{\lambda}^2) \approx -1.86$. One can see, that the value of the normalized threshold current $G_1$ does not depend on the angle $\phi$, thus, it is the solution of the problem of minimization of the functional $\Phi$. Since we disregard at this moment Gilbert damping, the found value of the current density *J* corresponds to the compensation of the radiation losses, and, as we see, this threshold value does not depend on the *i*-DMI. This feature will be explained below.

The last step is finding the generation frequency *W*. As it was pointed out, $\mathcal{F}(\phi) = 0$ if $W + \tilde{A}^2\cos^2\phi/4 = W_0 \approx 1.43$. The function $\mathcal{F}(\phi)$ close to this point can be expanded in a Taylor series as $\mathcal{F}(\phi) \approx C_\beta (\beta_1 - \beta_{1,0}) = C_\beta \left( \sqrt{W + \tilde{A}^2 \cos^2 \phi / 4} - \sqrt{W_0} \right)$ (one can directly verify, that $\mathcal{F}(\phi)$ is approximately linear in $\beta_1 = \sqrt{W + \tilde{A}^2 \cos^2 \phi / 4}$, but not in $W$). Using this expression in Eq. (13), one finds that the minimum of the functional $\Phi$ is achieved at $W = W_0 - \tilde{A}^2 / 4$ with the accuracy of $O(\tilde{A}^4)$, that is the solution we are searching for. Returning to the initial variables the generation frequency can be expressed as:

$$\omega = \omega_0 + 1.43 \omega_M \frac{\tilde{\lambda}^2}{R_{\text{eff}}^2} - \omega_M \frac{\tilde{D}^2}{4\tilde{\lambda}^2}. \tag{16}$$

The threshold current density is found after the addition of the Gilbert damping contribution. In the range of small values of the Gilbert damping (compared to the radiation losses) this contribution is simply equal to $\sigma J_G = \alpha_G \omega$,[14] because small damping does not change the spin wave profiles, and, consequently, radiation losses. In this case its role is simply to increase the threshold current to the value $\sigma J = \sigma J_0 + \alpha_G \omega$, so that the "negative damping" in the active area $\Gamma_- = \sigma J - \alpha_G \omega$ reaches the threshold value $\Gamma_{-,th} = \sigma J_0$. Thus, summarizing all the contributions, the threshold current density turns out to be :

$$\sigma J_{\text{th}} = 1.86 \omega_M \frac{\tilde{\lambda}^2}{R_{\text{eff}}^2} + \alpha_G \omega. \tag{17}$$

Equations (16) and (17) are the central results of the presented analytical model. In the limit of a zero *i*-DMI they are reduced to the ones, derived in Ref. 14, as it should be.

**(e) Analysis of the obtained equations**

According to Eq. (15) the presence of the *i*-DMI leads to a red shift of the generation frequency. This shift is *independent* of the geometry of the SHO active area, i.e. on the $R_{\text{eff}}$, and is equal to $\Delta \omega = -\omega_M \tilde{D}^2 / 4\tilde{\lambda}^2$. The reason of the frequency shift is clear – the *i*-DMI results in a decrease of the minimum frequency in the spectrum of spin waves. Indeed, the expression for the spin-wave spectrum Eq. (3) can be rewritten as:

$$\omega_k = \omega_0 + \omega_M \tilde{\lambda}^2 \left( \left( k_x + 2\tilde{D}/\tilde{\lambda}^2 \right)^2 + k_y^2 \right) - \omega_M \tilde{D}/4\tilde{\lambda}^2, \tag{18}$$

i.e. the spectrum is shifted in the $k_x$ direction, and is lowered by the value of $\Delta\omega = -\omega_M \tilde{D}^2/4\tilde{\lambda}^2$. The last value is *exactly the same* as the red shift of the generation frequency. This is absolutely natural, because the exchange interaction results in a certain offset of the generation frequency from the minimum frequency in the spectrum. This offset is the same for any *i*-DMI, because the structure of the spectrum remains the same except for the $k_x$-shift, which the exchange interaction is not sensitive to. Thus, one can expect, that the red shift of the generation frequency $\Delta\omega = -\omega_M \tilde{D}^2/4\tilde{\lambda}^2$ remains the same in all the *i*-DMI range, not only in the range of relatively small values. Our simulations below confirm this expectation. Also, it becomes clear, that in the one-dimensional case (nanowire along *x*-direction), the red shift is also given by the same expression $\Delta\omega = -\omega_M \tilde{D}^2/4\tilde{\lambda}^2$, as shown by the exact one-dimensional analytical model in Ref. 38.

Above we have also found that, in the absence of Gilbert damping, the generation threshold is independent of the *i*-DMI. In this case the threshold is determined by the compensation of the radiation losses $\Gamma_{\text{rad}}$. The radiation losses are proportional to the spin-wave group velocity given by Eq. (11), so the total radiation losses are obtained after the integration over $\phi_k$, and are proportional to $\Gamma_{\text{rad}} \sim \langle k \rangle$, where the averaged spin-wave wave number is given by Eq. (12). Substituting the expression for the generation frequency (Eq. (16) into Eq. (12)) one finds that, in the range of relatively small *i*-DMI, $\langle k \rangle \approx k_0 \left( 1 - \left( \tilde{D}/(2k_0 \tilde{\lambda}^2) \right)^4 / 4 \right)$, where $k_0 = \sqrt{1.43}/R_{\text{eff}}$. In the above presented model we have neglected the terms of the order of $\tilde{D}^4$. Thus, the radiation losses are independent of the *i*-DMI within the model, and, naturally, the obtained threshold current is also independent of the *i*-DMI. Expression for $\langle k \rangle$ gives, also, the range of the *i*-DMI, where model is valid: $\left(\tilde{D}/\left(2k_0\tilde{\lambda}^2\right)\right)^4/4 \ll 1$. Outside this range, one may expect a decrease of the threshold current since the averaged group velocity decreases. Moreover, if $|\tilde{D}| > 2k_0\tilde{\lambda}^2$ the spin waves in some direction become non-propagating (evanescent), since their wave vector becomes imaginary (see Eq. (10)). This feature was observed in simulations in Ref. [39]. However, to calculate the threshold dependence on this region analytically one should find a way to describe a general solution without approximation of the small values of *i*-DMI, which lies beyond the scope on this article.

## IV. Comparison with micromagnetic simulations and discussion

In this section, we compare predictions of the above presented analytical model with the results of our micromagnetic simulations. The geometry and parameters of our micromagnetic simulations are described in Sec. II, and the value of the bias magnetic field was 400 mT. In this case the parameters determined by means of the analytical model are equal to: FMR frequency $\omega_0 = 2\pi \times 7.81\,\text{GHz}$, effective exchange constant $\tilde{\lambda} = 5.64\,\text{nm}$, effective *i*-DMI parameter $\tilde{D} = D \times 0.62\,\text{nm}$, where $D$ is expressed in mJ/m$^2$. The effective radius of the active region is estimated from the difference of the generation frequency from the FMR frequency in the absence of the *i*-DMI. In the simulations we found $\omega_0 = 2\pi \times 7.8\,\text{GHz}$ and $\omega_{\text{gen}} = 2\pi \times 8.7\,\text{GHz}$, which, according to Eq. (15), results in the effective radius $R_{\text{eff}} = 42.2\,\text{nm}$. The effective radius is close to the half distance between the concentrator tips, as should be expected.

First, in Fig. 2 we compare analytical approximation Eqs. (8, 9) of the profile of excited spin-wave mode with the micromagnetic ones. One can clearly see, that spin wave profiles deviate from a purely cylindrical symmetry, and this deviation increases with the *i*-DMI t, as expected. The analytical approximation describes micromagnetic spin wave profiles reasonably well, and a weak deviation is related to the spatial distribution of the spin current, which is not of a perfect radial symmetry (see, e.g., supplementary materials in Ref. [38], as was assumed in the model).

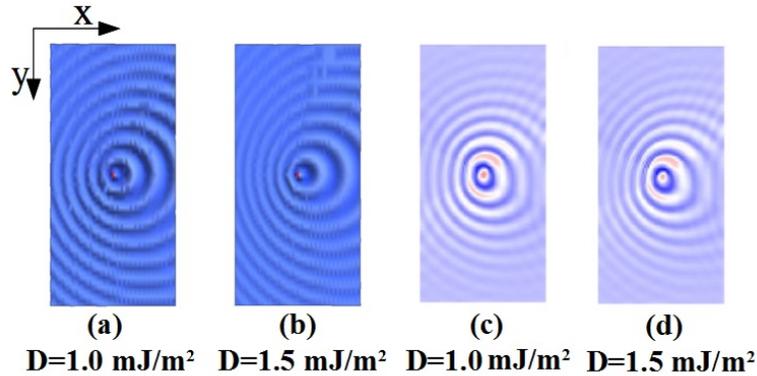

(a) D=1.0 mJ/m²  (b) D=1.5 mJ/m²  (c) D=1.0 mJ/m²  (d) D=1.5 mJ/m²

Fig. 2. Spatial profile of the excited spin-wave mode at different *i*-DMI strengths, (a) and (b) – theory (real part of the solution Eqs. (8, 9)), (c) and (d) – micromagnetic simulations. The rectangular cross section is 1500 nm × 3000 nm.

Quantitative comparison of the spin wave profiles can be made via the calculation of the angular dependence of the spin-wave wave number. Analytically, this dependence is given by Eq. (10), in which one should calculate the generation frequency using Eq. (16). Micromagnetic dependence was found by calculation of distances between the zeros directly from the time evolution of the spatial distribution of the magnetization. Spin-wave wave number monotonically increases when the azimuthal angle is varied from $\phi=0°$ (+x direction) to $\phi=180°$ (–x direction); at negative angles the dependence is symmetric, $k(-\phi)=k(\phi)$. The maximum difference of the wave numbers $k(180°)-k(0)$ is determined solely by the *i*-DMI strength, while the mean value, mainly, by the size of the active region. Again, we note quite a good description of the micromagnetic data by the analytical expression.

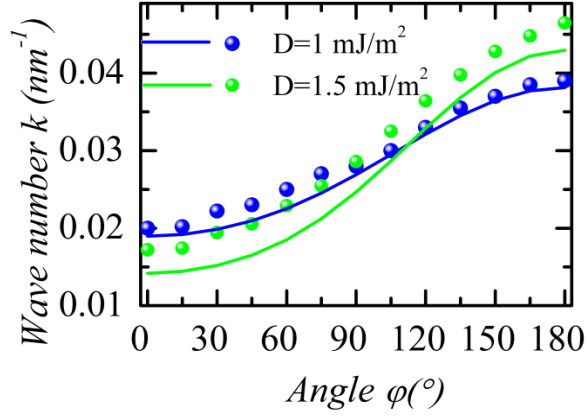

Fig. 3. Wave number of excited propagating spin-wave mode at different strength of *i*-DMI; symbols – micromagnetic simulation, lines – analytical expression (Eq. (10)).

Next, we look at the dependence of the generation frequency on the *i*-DMI, which is shown in Fig. 4(a). Simulated frequencies follow the predicted trend, and decrease with the *i*-DMI as $\Delta\omega = -\omega_M \tilde{D}^2 / 4\tilde{\lambda}^2$. It should be noted, that the equality of characteristic contributions of the *i*-DMI and non-uniform exchange interaction, which corresponds to the condition $|\tilde{D}| = 2k_0 \tilde{\lambda}^2$ (when the argument of the elliptic integral in Eq. (12) is equal to 1), in our case takes place at the *i*-DMI strength $D = 2.93$ mJ/m$^2$. Thus, the red shift of the generation frequency follows the same trend not only in the range of relatively small *i*-DMI values, but remains the same for a large *i*-DMI, as was predicted in Sec. III(e).

To prove additionally this feature, we analyzed the data of micromagnetic simulation in Ref. [39], where STO with an active area of exactly circular shape was studied. We use the data presented for the smallest bias current (3 mA), for which the nonlinear effects should be small. In that case, the characteristic value of the *i*-DMI, when its effect becomes the same as the effect of exchange interaction, is 0.85 mJ/m$^2$. As one can see from the inset in Fig. 4(a), the generation frequency follows the dependence of Eq. (16) in all the studied *i*-DMI range, including the range, where *i*-DMI becomes dominant ($D > 0.85$ mJ/m$^2$).

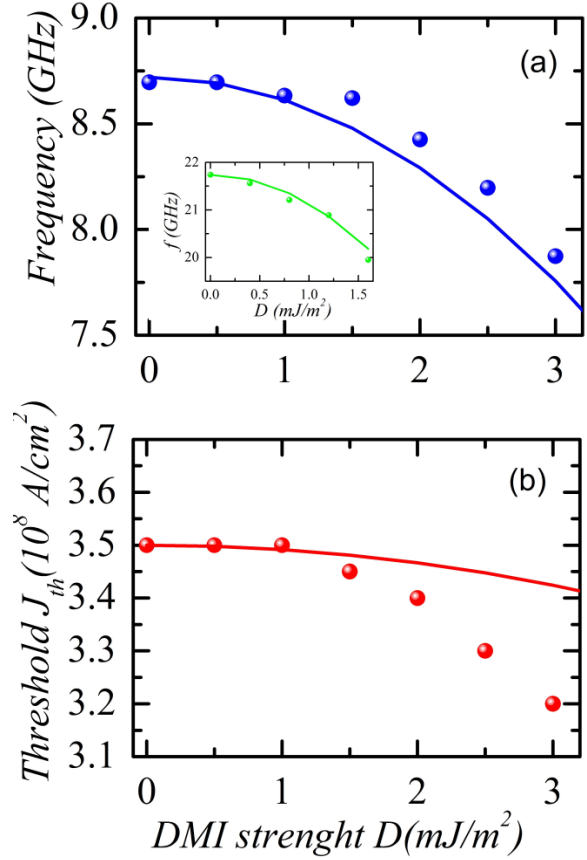

Fig 4. Dependences of the generation frequency (a) and threshold current density (b) on the *i*-DMI strength: symbols – micromagnetic data, lines – analytical model Eqs. (16) and (17), respectively. Inset in (a) shows the dependence of the generation frequency for the STO studied micromagnetically in Ref. [39]: points are the micromagnetic data retrieved from Fig. 2(a) in Ref. [39] at the bias current of 3 mA, line shows the result of the analytical model Eq. (16).

For the calculation of the threshold current density one needs the value of the spin-Hall efficiency $\sigma = \sigma_0 \sin\theta_M$. The theoretically calculated value is $\sigma_0 = 5.8 \times 10^{-3}\, \mathrm{m^2/(A \cdot s)}$. By determining the value of $\sigma_0$ from the matching of the calculated threshold by means of Eq. (17) in the absence of *i*-DMI and the micromagnetic data, we get a slightly higher value of $\sigma_0 = 6.6 \times 10^{-3}\, \mathrm{m^2/(A \cdot s)}$. This discrepancy is, mainly, attributed to a non-uniform spatial distribution of the current density, created by the concentrators. Below, we use the last value of the spin-Hall efficiency for analytical calculations of the threshold current.

According to Eq. (17), which is valid in the range of relatively small *i*-DMI, the threshold current weakly depends on the *i*-DMI, because only the Gilbert losses are dependent on the *i*-DMI due to *i*-DMI-induced red shift of the generation frequency, while the radiation losses don't depend on the *i*-DMI. In the range of relatively small *i*-DMI values ($D \leq 1.5 \text{mJ/m}^2$) our micromagnetic simulations confirm this prediction. However, when the strength of the *i*-DMI becomes comparable to the strength of the exchange interaction, we observed a decrease of the generation threshold current. As was pointed in Sec. IIIe, this decrease is related with a decrease of the averaged spin-wave group velocity, and, consequently, of the radiation losses.

Finally, we should note, that the presented theory is rigorously valid for the STOs with circular active region, while in the case of an SHO with concentrators one needs to use adjusting parameters: effective radius $R_{\text{eff}}$ and modified spin-Hall efficiency $\sigma$. To check if these parameters are set solely by the geometry of the concentrators we made simulations for different values of the bias magnetic field, which leads to a different magnetization angle, and compared these results with the corresponding curves calculated analytically . The *i*-DMI in this part of study is not taken into account, since the effects of the *i*-DMI on the generation frequency and threshold don't depend on the $R_{\text{eff}}$ (see Eqs. (16, 17)). As one can see from Fig. 5(a), the generation frequency has a constant offset from the FMR frequency, and is almost perfectly described by the analytical expression Eq. (16) with a constant $R_{\text{eff}}$ = 42.2 nm. The dependence of the threshold current density also agrees very well the numerically calculated one in all the bias field range, especially noting that the accuracy of the determination of the critical parameters in simulations are often not very high, because of the properties of numerical noise. Summarizing this part, we found, that the adjustuble parameters of the analytical model are determined by the current density distribution, and could be found from 1-2 reference points of micromagnetic simulations.

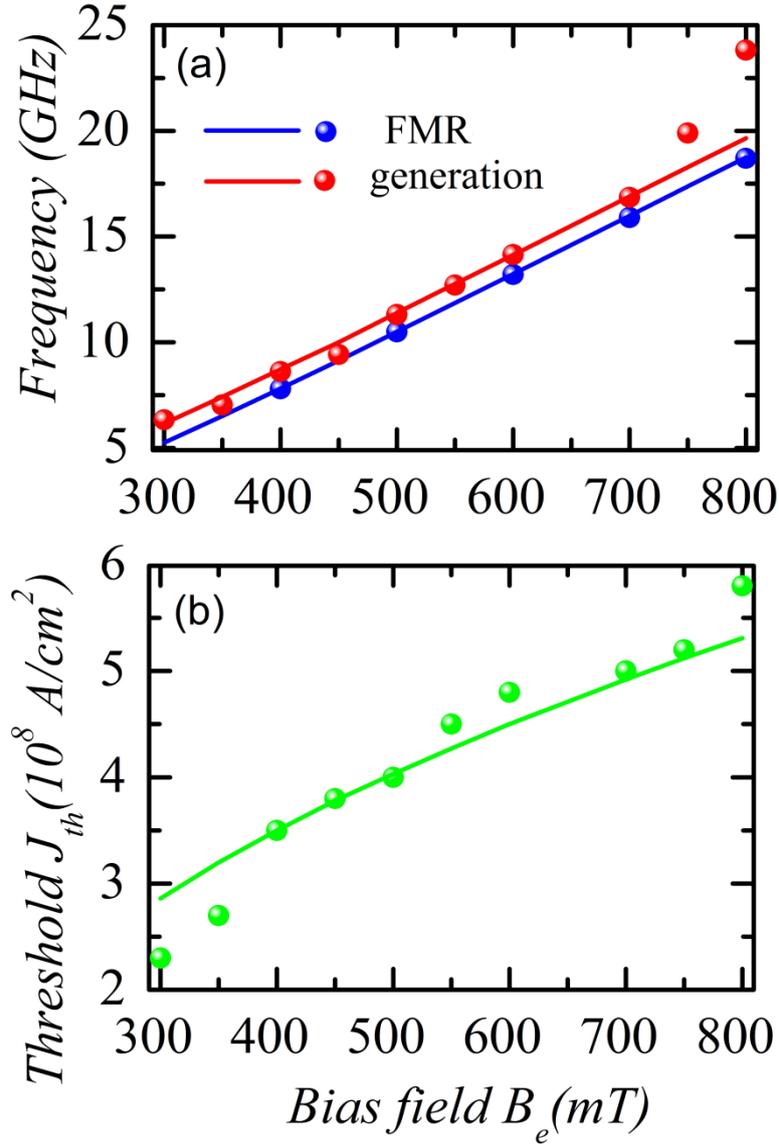

Fig.5. (a) Frequency of FMR and frequency of the excited spin waves at the threshold as functions of the bias magnetic filed. (b) Dependence of the threshold current on the bias magnetic field. Symbols – micromagnetic data, lines – analytical theory. Figures are plotted for the case of zero *i*-DMI.

## V. Conclusions

In summary, in this study we have proposed an analytical model for the description of the excitation of two-dimensional nonreciprocal spin waves in spin-torque and spin-Hall oscillators in the presence of *i*-DMI. In the range of weak and moderate *i*-DMI the analytical problem of the spin wave excitation is reduced to the eigenvalue problem for the generalized confluent Riemann equation. The profiles of the excited spin waves are described by a linear combination of a Laguerre's polynomial and a confluent hypergeometric function, and exhibit nonreciprocal behavior with the angular dependence of the spin-wave wave number.

It is shown that the frequency of the excited spin waves at the threshold exhibit a quadratic red shift with the increase of the *i*-DMI strength. This shift is a direct consequence of the lowering of the spin-wave spectrum bottom in the presence of the *i*-DMI. Therefore, this shift is proportional to the ratio between the characteristic *i*-DMI length and the exchange length, and could be expressed by the same functional dependence in all the studied *i*-DMI range, including the range , where *i*-DMI makes a dominant contribution to the properties of the excited spin waves.

At the same time, the averaged spin-wave wave number and spin-wave group velocity are almost independent of the *i*-DMI in the range of small and moderate *i*-DMI. Consequently, the radiation losses remain the same, and the *i*-DMI affects the excitation threshold current only via its weak influence on the Gilbert losses, which are proportional to the generation frequency. However, when the effect of the *i*-DMI becomes comparable or greater than that of the exchange interaction, we observed a decrease of the generation threshold, which is attributed to the decrease of the averaged group velocity.

## Acknowledgements

This work was supported by the executive programme of scientific and technological cooperation between Italy and China for the years 2016–2018 (code CN16GR09) title "Nanoscale broadband spin-transfer-torque microwave detector" funded by Ministero degli Affari Esteri e della Cooperazione Internazionale. This work was also supported in part by the Grants Nos. EFMA-1641989 and ECCS-1708982 from the NSF of the USA, and by the DARPA M3IC Grant under the Contract No. W911-17-C-0031. R.Z. acknowledges the support by National Group of Mathematical Physics (GNFM-INdAM). R.V. acknowledges support from Ministry of Education and Science of Ukraine (project 0118U065874).